\documentclass[preprint,12pt]{revtex4-1}
\usepackage{amsmath}
\usepackage{amsfonts}
\usepackage{amssymb}
\usepackage{graphicx}
\usepackage{bm}
\begin{document}
\title{Quantum dynamics of the harmonic oscillator on a cylinder}
\author{K. Kowalski and J. Rembieli\'nski}
\affiliation{Department of Theoretical Physics, University
of \L\'od\'z, ul.\ Pomorska 149/153, 90-236 \L\'od\'z,
Poland}
\begin{abstract}
Evolution of coherent states is considered for a particle confined to
a cylinder moving in a harmonic oscillator potential.  Because of the
discontinuous changes as time goes by of the phase representing the
position of a particle on a parallel (circle) the trajectory pattern
of quantum averages specifying coordinates on a cylinder is very complex
and in some aspects resembles chaotic one.
\end{abstract}
\maketitle
\section*{}
Despite the fact that the helical motion of a charged particle in a uniform
magnetic field is described in many textbooks the theory of quantization on
a cylinder is far from complete.  For example, the coherent states for the
quantum mechanics on a cylinder were not, to the best of our knowledge, discussed
in the literature.  We point out that coherent states for a charged particle
in a magnetic field are obtained by reduction of the dynamics to the transversal
motion \cite{1}.  However, it is clear that in general such reduction cannot be 
applied.  An example is a particle on a cylinder with radial magnetic field
analyzed in \cite{2}.  In this work we study the dynamics of the harmonic 
oscillator on a cylinder.  More precisely, we discuss the time development of the
wave packets when the initial condition is a coherent state.  Since the coherent
states for a cylinder are not stable but do not spread as in the case of the
motion in a plane, the dynamics is nontrivial.  In particular, the geometry of 
jump points related to discontinuity of the phase representing the position
of a particle on a parallel is similar to the Poincar\'e section of chaotic
trajectories of nonlinear dynamical systems.

We first summarize the basic facts about the quantum mechanics on a cylinder
\begin{eqnarray}
x_1 &=& \cos\varphi,\nonumber\\
x_2 &=& \sin\varphi,\\
x_3 &=& l,\nonumber
\end{eqnarray}
where $\varphi\in[0,2\pi)$ specifies the position of a particle on a
parallel and $l\in(-\infty,\infty)$ is the coordinate of a particle on a
meridian (generator).  On taking into account the topological equivalence of
the infinite circular cylinder (1) with the product of the circle and real line
$S^1\times\mathbb R$, we arrive at the following algebra adequate for the study
of the motion on a cylinder:
\begin{equation}
\begin{split}
[\hat J,U]&=U,\qquad [\hat J,U^\dagger]=-U^\dagger,\qquad [\hat
l,\hat p_l]={\rm i}I,\\
[\hat J,\hat l]&= [\hat J,\hat p_l]=[U,\hat l]=[U,\hat p_l]=[U^\dagger,\hat l]
=[U^\dagger,\hat p_l]=0,
\end{split}
\end{equation}
where $\hat J$ is the angular momentum operator, $U=e^{{\rm i}\hat\varphi}$ is 
the unitary operator representing the position of a quantum particle on a (unit)
circle \cite{3},  $\hat l$ is the position operator on a generator of a cylinder,
$\hat p_l$ is the corresponding momentum, and we set $\hbar=1$.  The operators
act in the Hilbert space $L^2(S^1\times\mathbb R)$ for the quantum mechanics on 
a cylinder specified by the scalar product
\begin{equation}
\langle f|g\rangle =
\frac{1}{2\pi}\int_0^{2\pi}d\varphi\int_{-\infty}^{\infty}
dl\,f^*(\varphi,l)g(\varphi,l).
\end{equation}
In the following we restrict to the case of integer eigenvalues of the operator
$\hat J$, then the functions $f(\varphi,l)$ are periodic functions of $\varphi$
with period $2\pi$ \cite{3}.  The operators act in the representation (3) as 
follows
\begin{equation}
\begin{split}
{\hat J}f(\varphi,l) &= -{\rm
i}\frac{\partial}{\partial\varphi}f(\varphi,l),\qquad
Uf(\varphi,l)=e^{{\rm i}\varphi}f(\varphi,l),\\
{\hat p_l}f(\varphi,l) &= -{\rm i}\frac{\partial}{\partial l}
f(\varphi,l),\qquad{\hat l}f(\varphi,l)=lf(\varphi,l).
\end{split}
\end{equation} 

Consider now the coherent states for a particle on a cylinder.  The topological
equivalence of a cylinder and the product $S^1\times\mathbb R$ indicates that the
coherent states are common eigenvectors of the commuting operators: $X=
e^{-{\hat J}+\frac{1}{2}}U$ defining via an eigenvalue equation the coherent states
for the quantum mechanics on a circle \cite{3} and the annihilation operator 
$a=\frac{1}{\sqrt{2}}({\hat l}+{\rm i}{\hat p_l})$ used for the definition of 
the standard coherent states for a particle on a line, that is
\begin{subequations}
\begin{eqnarray}
Xf_{\xi,z}(\varphi,l)&=&\xi f_{\xi,z}(\varphi,l),\\
af_{\xi,z}(\varphi,l)&=& zf_{\xi,z}(\varphi,l),
\end{eqnarray} 
\end{subequations}
where $\xi=e^{-J+{\rm i}\alpha}$, where $J$ is the classical orbital momentum and
$\alpha$ is the classical angle parametrizing the classical phase space for a
particle on a circle \cite{3} and $z=\frac{1}{\sqrt{2}}(q+{\rm i}p)$.  Clearly,
the coherent state $f_{\xi,z}(\varphi,l)$ is simply the product of the coherent
states for the quantum mechanics on a circle and a real line in the coordinate
representation.  Hence, using the formulae for a particle on a circle
\cite{4} and well-known relations for the standard coherent states we find
\begin{equation}
f_{\xi,z}(\varphi,l) =\pi^{-1/4}\theta_3\left(\frac{1}{2\pi}
(\varphi-\alpha-{\rm i}J)\bigg\vert\frac{{\rm i}}
{2\pi}\right)\exp[-\hbox{$\scriptstyle\frac{1}{2}$}(l-q)^2
+ {\rm i}p(l-\hbox{$\scriptstyle\frac{1}{2}$}q)],
\end{equation}
where $\theta_3(v|\tau)$ is the Jacobi theta function \cite{5}. 

We now discuss the harmonic oscillator on a cylinder.  We begin by recalling
that the Hamiltonian for a classical particle with unit mass confined to a surface
of a cylinder (1) moving in a harmonic oscillator potential is given by
\begin{equation}
H = \frac{p_l^2}{2} + \frac{J^2}{2} + \frac{\omega^2}{2}l^2,
\end{equation}
where $p_l={\dot l}$ is the momentum corresponding to the motion in 
the meridian and $J={\dot\varphi}$ is the conserved angular momentum.
We point out that the potential $\frac{\omega^2}{2}l^2$ differs only
by a constant from $\frac{\omega^2}{2}{\bm x}^2~=~\frac{\omega^2}{2}
(x_1^2+x_2^2+x_3^2)$.  Of course, besides the circular motion in the $x_1$ and
$x_2$ plane and an equilibrium point, the solution describes the superposition of
the uniform circular motion
\begin{equation}
\varphi = \varphi_0 +Jt,
\end{equation}
with the angular velocity $\omega_J=J$, and the harmonic oscillations along the 
meridian with frequency $\omega$
\begin{equation}
l = l_0\cos\omega t + \frac{p_{l0}}{\omega}\sin\omega t,\qquad
p_l = p_{l0}\cos\omega t - \omega l_0\sin\omega t.
\end{equation}
In the case of commensurable $\omega$ and $\omega_J$ the motion is periodic and 
the trajectory is the closed curve on the cylinder with the upper and lower bound
$l=\pm\sqrt{l_0^2+\left(\frac{p_{l0}}{\omega}\right)^2}$, respectively, otherwise 
the motion is quasiperiodic and the trajectory densely fills the strip 
$|l|\le\sqrt{l_0^2+\left(\frac{p_{l0}}{\omega}\right)^2}$, $0\le\varphi<2\pi$.

In quantum mechanics the dynamics of the harmonic oscillator on a cylinder 
defined by the classical Hamiltonian (7) is described by the Schr\"odinger
equation
\begin{equation}
{\rm i}\frac{\partial f(\varphi,l;t)}{\partial t}={\hat H}f(\varphi,l;t),
\qquad f(\varphi,l;0)=f_0(\varphi,l),
\end{equation}
where the quantum Hamiltonian is
\begin{equation}
{\hat H} = \frac{{\hat p_l}^2}{2} + \frac{{\hat J}^2}{2} + 
\frac{\omega^2}{2}{\hat l}^2.
\end{equation}
As is well known the Hamiltonian $\hat H$ can be written in the form
\begin{equation}
\hat H = \frac{{\hat J}^2}{2} + \omega\left(N_\omega + 
\frac{1}{2}\right),
\end{equation}
where $N_\omega=a_\omega^\dagger a_\omega$ is the number operator
expressed by means of the Bose annihilation operators
\begin{equation}
a_\omega = \sqrt{\frac{\omega}{2}}\left({\hat l}+\frac{{\rm
i}}{\omega}{\hat p_l}\right).
\end{equation}
The coherent states of the harmonic oscillator with the Hamiltonian
(11) can be immediately obtained by the formal generalization of coherent
states defined by (5) relying on replacement of (5.b) with
\begin{equation}
a_\omega f_{\xi,z;\omega}(\varphi,l) = zf_{\xi,z;\omega}(\varphi,l),
\end{equation}
where $z=\sqrt{\frac{\omega}{2}}\left(q+\frac{{\rm i}}{\omega}p\right)$.
Hence, using the coordinate representation of the standard coherent 
states of the harmonic oscillator we find that the coherent states
$f_{\xi,z;\omega}(\varphi,l)$ are given by
\begin{equation}
f_{\xi,z;\omega}(\varphi,l)=\left(\frac{\omega}{\pi}\right)^\frac{1}{4}
\theta_3\left(\frac{1}{2\pi}(\varphi-\alpha-{\rm i}J)\bigg\vert\frac{{\rm i}}
{2\pi}\right)\exp\left[-\frac{\omega}{2}(l-q)^2
+ {\rm i}p(l-\hbox{$\scriptstyle\frac{1}{2}$}
q)\right].
\end{equation}
Suppose now that the initial state $f_0(\varphi,l)$ in (10) is the coherent 
state (15).  On taking into account that $\hat J$ and $\hat N_\omega$ commute, 
making use of the formula for the free evolution of the coherent states on
a circle \cite{4} and well-known relations describing stability of the standard 
coherent states with respect to the Hamiltonian of the harmonic-oscillator, we 
arrive at the following solution to the Schr\"odinger equation (10):
\begin{eqnarray}
&&f_{\xi,z;\omega}(\varphi,l;t) = e^{-{\rm i}t{\hat J}^2/2}
e^{-{\rm i}t\omega\left(N_\omega+\frac{1}{2}\right)}
f_{\xi,z,\omega}(\varphi,l)\\\nonumber
&&=\left(\frac{\omega}{\pi}\right)^\frac{1}{4}\theta_3\left(\frac{1}{2\pi}
(\varphi-\alpha-{\rm i}J)\bigg\vert\frac{1}{2\pi}({\rm i}-t)\right)
e^{-{\rm i}t\omega/2}\exp\left[-\frac{\omega}{2}l^2\right.\\\nonumber
&&\left.{}-\frac{1}{2}\omega q^2e^{-{\rm i}\omega t}\cos\omega t
-\frac{{\rm i}}{2\omega}
p^2e^{-{\rm i}\omega t}\sin\omega  t-\frac{1}{2}
{\rm i}qpe^{-2{\rm i}\omega t} 
+\omega\left(q+\frac{{\rm i}}{\omega}p\right) 
e^{-{\rm i}\omega t}l\right].
\end{eqnarray}
Since the Jacobi theta function $\theta_3(v,\tau)$ is a periodic
function of $\tau$ with period $T=2$ \cite{5}, so in view of (16)
the frequency of circular motion is equal to $\frac{1}{2}$, therefore
the function $f_{\xi,z;\omega}(\varphi,l;t)$ is periodic function of
time for rational $\omega$ and quasiperiodic for irrational $\omega$.
Furthermore, from (16) it follows that the probability density for the 
coordinates in the normalized coherent state is
\begin{eqnarray}
&&p_{\xi,z}(\varphi,l;t)=\frac{|f_{\xi,z;\omega}(\varphi,l;t)|^2}
{\|f_{\xi,z;\omega}\|^2}\\\nonumber
&&=\sqrt{\frac{\omega}{\pi}}\frac{\big\vert\theta_3\left(\frac{1}{2\pi}
(\varphi-\alpha-{\rm i}J)\big\vert
\frac{1}{2\pi}({\rm i}-t)\right)\big\vert^2}
{\theta_3\left(\frac{{\rm i}J}{\pi}\big\vert\frac{{\rm i}}
{\pi}\right)}\exp\left[-\omega\left(l-q\cos\omega t-\frac{p}{\omega}
\sin\omega t\right)^2\right].
\end{eqnarray}
As with (16) the probability density is periodic (quasiperiodic) for
rational (irrational) $\omega$.  We point out that the coherent
states for a particle on a circle are not stable with respect to the
free evolution generated by the Hamiltonian ${\hat J}^2/2$.
However, in opposition to the case of the standard coherent states
for a particle on a real line, the wave packets referring to the
angular part of the function (16) do not spread but oscillate during
the course of time.  As a result of oscillations of the corresponding 
probability density i.e.\ the angular part of (17), an interesting 
phenomenon occurs which can be regarded as quantum jumps on a circle 
described in \cite{4}.  We recall that the probability 
density has at $t=t_*=(2k+1)\pi$, where $k$ is integer, two identical 
maxima.  Therefore at $t=t_*$ a particle can be detected with equal 
(maximal) probability at two different points on a circle.  As a 
consequence of such behavior of the probability density there is  a
discontinuity in the angle that on the quantum level is identified
with ${\rm Arg}\langle U(t)\rangle$, where $U(t)~=~e^{{\rm i}t{\hat J}^2/2}
Ue^{-{\rm i}t{\hat J}^2/2}$, related to the jumps of the phase by
$\pi$.  Such dicontinuity takes place in the discussed case of integer
eigenvalues of $\hat J$ only for $J$ integer.  Now, in the light of 
the observations concerning a circle it is clear that in the discussed 
case of the wave packets (16) we deal with quantum jumps on a cylinder.  
The jumps occur at $t=t_*=(2k+1)\pi$, where $k$ is integer, when the
probalility density (17) has two identical maxima (see Fig.\ 1),
\begin{figure*}
\centering
\begin{tabular}{c}
\includegraphics[scale=1]{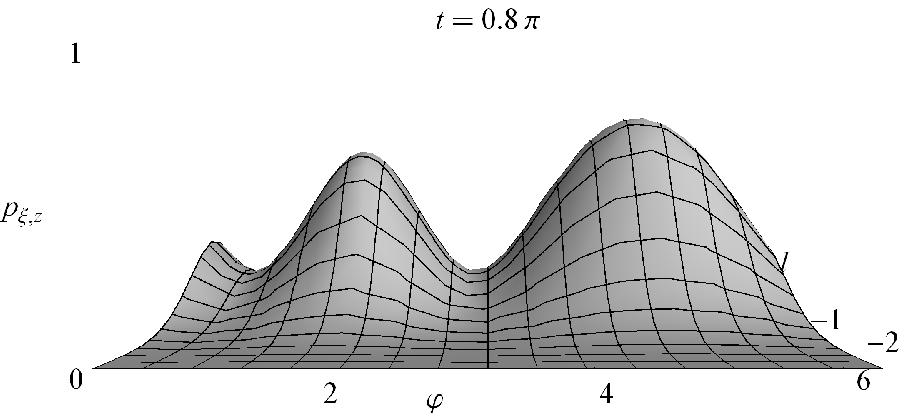}\\
\includegraphics[scale=1]{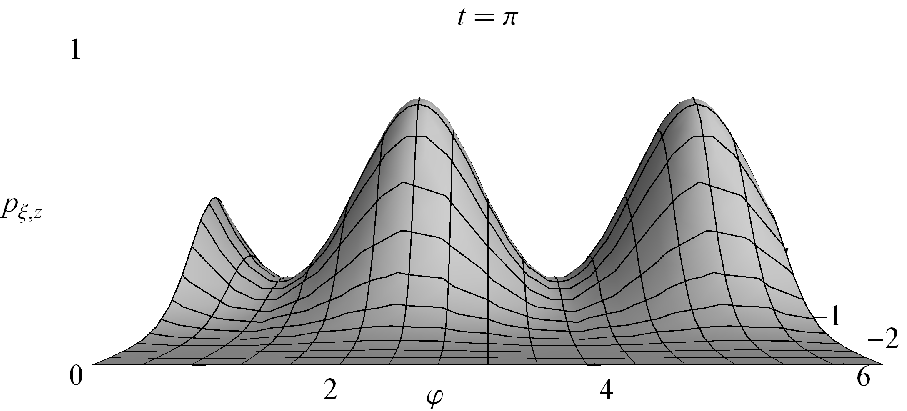}\\
\includegraphics[scale=1]{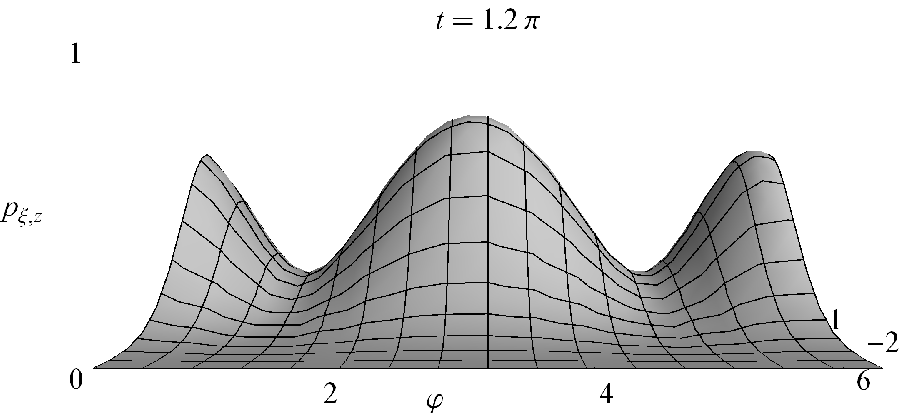}
\end{tabular}
\caption{The front-view of quantum probability density (17), where
$\omega =1$, $\alpha=0.75\pi$, $J=1$, $q=-0.7$, and $p=0.2$.  For 
$t=\pi$ the two maxima are identical, so a particle can be detected 
with equal probability on two different points on a cylinder.}
\end{figure*}
and they are related to the discontinuity in the angle ${\rm Arg}\langle 
U(t)\rangle_{\xi,z}$, where $\langle U(t)\rangle_{\xi,z}$ is the
expectation value of the operator $U(t)$ in the normalized coherent
state $f_{\xi,z;\omega}$ given by \cite{4}
\begin{equation}
\langle U(t)\rangle_{\xi,z}=e^{-1/4}e^{{\rm i}\alpha}\
\frac{\theta_2\left(\frac{t}{2\pi}-\frac{{\rm i}J}{\pi})
\big\vert\frac{{\rm i}}{\pi}\right)}{\theta_3\left(\frac{{\rm i}J}
{\pi}\big\vert\frac{{\rm i}}{\pi}\right)}.
\end{equation}
Such discontinuity takes place only for $J$ integer and
is connected with the jumps of phase by $\pi$ corresponding to the
transformation $(x_1,x_2)\to(-x_1,-x_2)$ of the coordinates of a
particle on a cylinder.  We study the quantum dynamics of the harmonic 
oscillator on a cylinder by means of expectation values of operators 
representing the angle and position on a meridian (generator).  More 
precisely, we set in (1)
\begin{equation}
\varphi = {\rm Arg}\langle U(t)\rangle_{\xi,z},\qquad 
l = \langle{\hat l}(t)\rangle_{\xi,z} = q\cos\omega t
+\frac{p}{\omega}\sin\omega t,
\end{equation}
where $\langle U(t)\rangle_{\xi,z}$ is given by (18), and 
${\hat l}(t)~=~e^{{\rm i}t\omega\left(N_\omega +\frac{1}{2}\right)}
{\hat l}e^{{-\rm i}t\omega\left(N_\omega +\frac{1}{2}\right)}$.  We
point out that $\theta_2(v|\tau)$ is a periodic function of $v$ with
period $T=2$ \cite{5}, so $\varphi$ given by (18) and (19) is periodic
with period $4\pi$ and thus the frequency of circular motion is $1/2$.
We conclude that the trajectory on a cylinder given by (1) and
(19) is periodic for rational $\omega$ and quasiperiodic for
irrational $\omega$.  Interestingly, as follows from computer simulations
the jump points concentrate around the generator corresponding to the 
angle $\varphi=\alpha-\frac{\pi}{2}$, where $\alpha$ is the parameter 
labelling the coherent state refering to the classical angle.  Since 
both the classical and quantum dynamics of the harmonic oscillator on 
a cylinder given by (8), (9) and (19), respectively is quasiperiodic for 
irrational frequency $\omega$, no wonder that the set of jump points 
turns out to be very complicated for irrational $\omega$ that is 
illustrated in Fig.\ 2.  We point out that the geometry of jump points
presented in Fig.\ 2 resembles the Poincar\'e section of chaotic
trajectories of nonlinear dynamical systems.  The difference of the 
pattern referring to the quasiperiodic motion with irrational $\omega$ 
and periodic motion with rational $\omega$ is also remarkable.
\begin{figure*}
\centering
\begin{tabular}{c}
\includegraphics[scale=1]{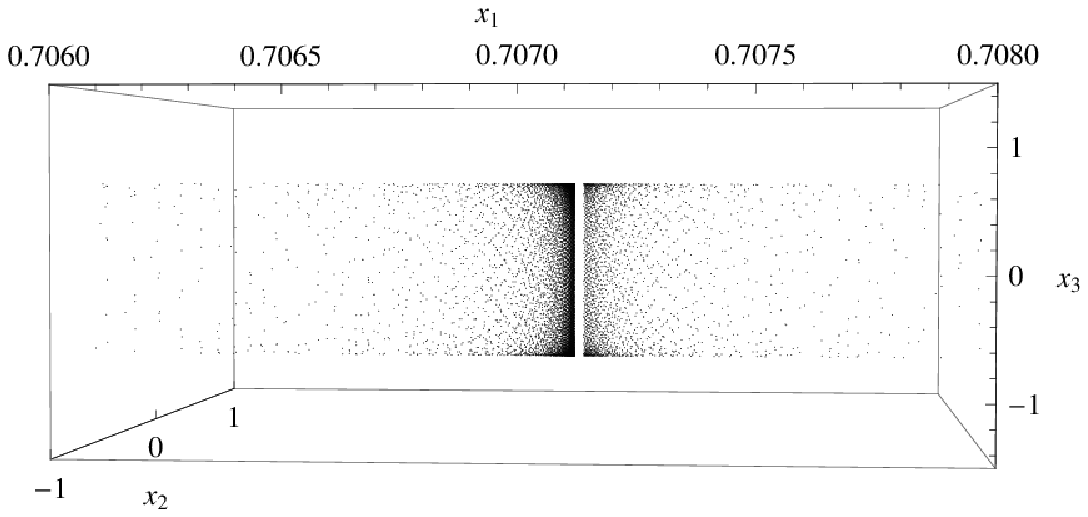}\\
\includegraphics[scale=1]{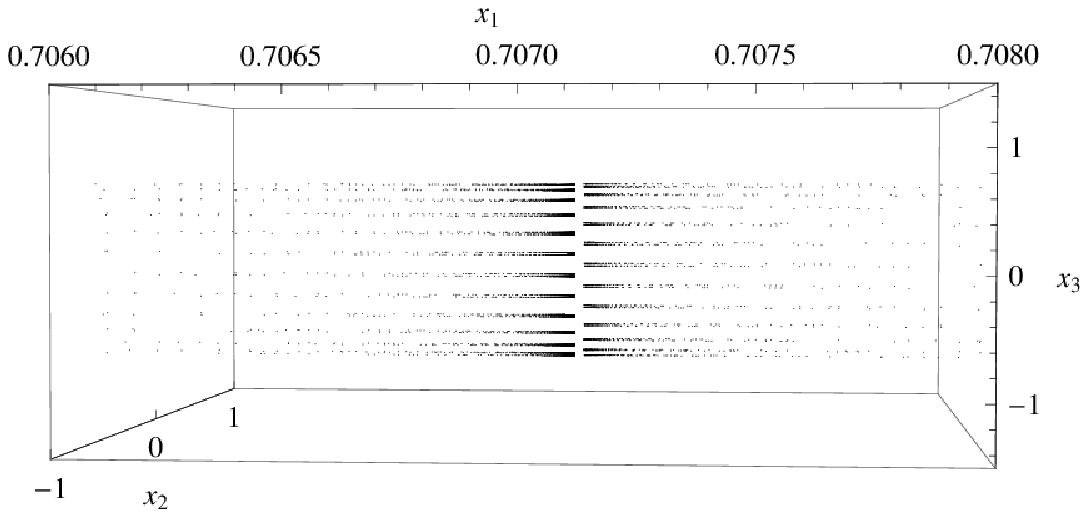}
\end{tabular}
\caption{Top: fragment of the surface of the cylinder with
jump points, where $\omega=\frac{1}{2}(1+\sqrt{5})\approx1.61803$, 
that is $\omega$ is the golden ratio --- the most irrational number 
of all irrational numbers.  The remaining parameters are the same as
in Fig.\ 1.  Bottom: the jump points in the case with
rational $\omega=1.62$ approximating the golden ratio, that is the
frequency for the top figure.  Despite the fact that the relative
error of the approximation of the golden ratio by $\omega=1.62$ is of
order $0.1\%$, the difference of the pattern from the top and bottom
figure referring to the classical quasiperiodic and periodic motion,
respectively is noticable.}
\end{figure*}

Concluding, a quantum particle on a cylinder in a coherent state moving
in a harmonic oscillator potential shows exotic behaviour which can be
interpreted as pseudo-stochastic quantum jumps.  It seems that besides
the quantum mechanics of constrained systems, the results concerning
the dynamics of the harmonic oscillator on a cylinder would be also of 
importance for the theory of quantization of a quasiperiodic motion.
Finally, the provided example of nontrivial dynamics with discontinuous
trajectories would be also of interest in the theory of dynamical systems.
\section*{Acknowledgements}
This work was supported by the grant N202 205738 from the National
Science Centre.


\begin{thebibliography}{}
\bibitem{1}K. Kowalski, J. Rembieli\'nski, J. Phys. A 38 (2005) 8247.
\bibitem{2}C. Chryssomalakos, A. Franco, A. Reyes-Coronado, Eur. J. Phys.
25 (2004) 489.
\bibitem{3}K. Kowalski, J. Rembieli\'nski, L.C. Papaloucas, J.
Phys. A 29 (1996) 4149.
\bibitem{4}K. Kowalski, J. Rembieli\'nski, Phys. Lett. A 293 (2002) 109.
\bibitem{5}G.A. Korn and T.M. Korn, Mathematical Handbook, McGraw-Hill,
New York, 1968.
\end{thebibliography}
\end{document}